# PRELIMINARY ASSESSMENT OF AN INTEGRATED SMOS AND MODIS APPLICATION FOR GLOBAL AGRICULTURAL DROUGHT MONITORING

*N. Sánchez[1], A. González-Zamora[1], J. Martínez-Fernández[1], M. Piles[2], M. Pablos[1], Brian Wardlow[3], Tsegaye Tadesse[4], Mark Svoboda[4]*

[1] Instituto Hispano Luso de Investigaciones Agrarias (CIALE), University of Salamanca. Duero, 12. 37185. Villamayor. Spain. (nilda, aglezzamora, jmf, mpablos)@usal.es
[2] Image Processing Lab (IPL), Universitat de Valencia, Spain, maria.piles@uv.es
[3] Center for Advanced Land Management Information Technologies (CALMIT), University of Nebraska-Lincoln, bwardlow2@unl.edu
[4] National Drought Mitigation Center (NDMC), University of Nebraska-Lincoln, (ttadesse2, msvoboda2) @unl.edu

**ABSTRACT**

An application of the Soil Moisture Agricultural Drought Index (SMADI) at the global scale is presented. The index integrates surface soil moisture from the SMOS mission with land surface temperature (LST) and Normalized Difference Vegetation Index (NDVI) from MODIS and allows for global drought monitoring at medium spatial scales (0.05 deg).. Biweekly maps of SMADI were obtained from year 2010 to 2015 over all agricultural areas on Earth. The SMADI time-series were compared with state-of-the-art drought indices over the Iberian Peninsula. Results show a good agreement between SMADI and the Crop Moisture Index (CMI) retrieved at five weather stations (with correlation coefficient, R from -0.64 to -0.79) and the Soil Water Deficit Index (SWDI) at the Soil Moisture Measurement Stations Network of the University of Salamanca (REMEDHUS) (R=-0.83). Some preliminary tests were also made over the continental United States using the Vegetation Drought Response Index (VegDRI), with very encouraging results regarding the spatial occurrence of droughts during summer seasons. Additionally, SMADI allowed to identify distinctive patterns of regional drought over the Indian Peninsula in spring of 2012. Overall results support the use of SMADI for monitoring agricultural drought events world-wide.

## 1. INTRODUCTION

Remote sensing-based drought indices are emerging as an effective tool for large-scale drought monitoring in comparison with ground-based indices, which are difficult to gather and generalize at the global scale. The ground-based indices are usually derived from meteorological variables such as precipitation and temperature, and its application at local or regional scales depends mainly on the density and distribution of the ground station networks [1]. By contrast, remote sensing indices are appropriate for large-scale applications and can integrate agricultural drought indicators as soil moisture or vegetation information [2, 3]. In fact, since agricultural drought is closely related to soil moisture deficit, the development of microwave missions dedicated to measuring soil moisture (i.e., SMOS, Soil Moisture and Ocean Salinity and SMAP, Soil Moisture Active Passive) have positioned this essential variable as a key tool to identify and, in some cases, assess agricultural drought [2, 4, 5].

Drought indices are generally used to provide a quantitative analysis of the severity, location, timing and duration of drought events [6]. The Soil Moisture Agricultural Drought Index (SMADI) [7] is used in this study. It was designed to assess the agricultural drought at the global scale integrating the response of the soil, the vegetation and close-atmosphere layers from satellite products. SMADI embeds any remote sensing dataset of land surface temperature (LST), vegetation indices (e.g., the Normalized Difference Vegetation Index, NDVI) and surface soil moisture (SSM). SMADI is scalable in space and time, and allows integrating any remote sensing product providing LST, NDVI and SSM information (after some pre-treatment required for data integration). In this work, global daily products of LST and NDVI from the Moderate Resolution Imaging Spectro-radiometer (MODIS) at 0.05° and SMOS SSM were integrated. The resulting time-series of SMADI were compared with other agricultural drought indicators at a suite of weather stations in Spain, and the spatial patterns were compared to the Vegetation Drought Response Index (VegDRI) over the United States. Results show that SMOS/MODIS SMADI allows identifying and monitoring the evolution of drought occurrences in global agro-ecosystems.

## 2. DATA AND METHODS

### 2.1. SMADI rationale

SMADI is based on the inverse relationship between the LST and vegetation status, which are, in turn, related to the soil moisture content [8, 9]. The LST and NDVI are integrated in the index using the normalized form between the maximum and minimum time range for each pixel, the so-called temperature and vegetation condition indices (MTCI [3] and VCI [10]), respectively:

$$MTCI = \frac{(LST_i - LST_{min})}{(LST_{max} - LST_{min})} \quad (1)$$

$$VCI = \frac{(NDVI_i - NDVI_{min})}{(NDVI_{max} - NDVI_{min})} \quad (2)$$

Similarly, SSM is integrated in SMADI using a normalization with respect to the maximum and minimum time range values for each pixel, the so-called soil moisture condition index (SMCI):

$$SMCI = \frac{(SSM_{max} - SSM_i)}{(SSM_{max} - SSM_{min})} \quad (3)$$

Finally, SMADI is computed as the slope of the MTCI and VCI, and includes the SMCI as a multiplicative factor, as follows:

$$SMADI_i = SMCI_i \frac{MTCI_i}{VCI_{i+1}} \quad (4)$$

where $i$ corresponds to a given biweekly period. Note that the VCI selected for a given $i$ corresponds to the ensuing fourteen-day period, in order to consider the time lag between the plant response and the soil moisture conditions [11].

### 2.2. Remote sensing inputs for SMADI

SMADI was previously tested at 500 m of spatial resolution over the Iberian Peninsula with very satisfactory results [7]. The main objective of this experiment is to assess the suitability of SMADI at the global scale. To do so, the following global MODIS imagery was selected: the daily MODIS/Terra LST (MOD11C1 v.6) (only daytime was selected, owing the previous results in [7]), and the daily reflectance (MOD09CMG v.6), from which the NDVI was retrieved. Both products were provided on a 0.05° latitude/longitude global grid. Later, they were time-averaged into biweekly series using the 14 antecedent days.

Regarding the SSM, the daily SMOS L2 v.620 soil moisture retrieved at 15 km grid was selected. Then, the SSM was transformed into the same global 0.05° regular grid and biweekly timing. The study period spans from June, 2010, when SMOS products are available after its commissioning phase, to December, 2015.

All input data were firstly masked using a land-use map at 0.05° coming from the MODIS/Terra Land Cover Types map of 2012 (MOD12C1), including classes of grassland, rainfed agriculture and cropland/natural vegetation mosaic as agriculture regions for SMADI. This clustering aims to ensure the study is focused in crop areas where the sole supply of water comes from precipitation and the water availability strongly limits vegetation growth.

### 2.3. Ground observations for CMI and SWDI

Ground meteorological data were acquired from the Spanish Meteorological Agency (AEMet) at five automatic weather stations distributed over the centre of the Iberian Peninsula. Historical series of temperature and precipitation, together with soil properties, were used to calculate the Crop Moisture Index (CMI) [12] at this stations. In addition, the Soil Water Deficit Index (SWDI) [2] was calculated based on the soil moisture observations from the Soil Moisture Measurement Stations Network of the University of Salamanca (REMEDHUS) [13]. Both CMI and SWDI are also agricultural drought indices and their time-series were compared to the SMADI series at the given locations. Note the interpretation of their magnitude are though opposite to SMADI's, i.e., the smaller the indices, the higher the drought.

### 2.4. VegDRI

VegDRI is a composite, agricultural drought indicator that integrates satellite, climate, and biophysical data to characterize the relative severity of drought conditions. VegDRI models were developed sing an empirical, regression tree approach based on the analysis of a 20+ year historical record of satellite (percent annual seasonal greenness and start of season anomaly), climate (Standardized Precipitation Index (SPI) and calibrated Palmer Drought Severity Index (scPDSI) and biophysical (land use/land cover type, soils, elevation, and ecoregion type) variables. Additional technical details about the VegDRI are presented by [14], [15] and [7]. VegDRI is an operational monitoring tool over the continental United States supported by the National Drought Mitigation Center (NDMC) and U.S. Geological Survey with 1-km resolution VegDRI maps being produced on a bi-weekly time step with the time series of bi-weekly maps dating back to 1989. Historical bi-weekly VegDRI data were acquired from the NDMC for this study.

## 3. RESULTS AND DISCUSSION

Series of 145 biweekly global maps of SMADI were retrieved during the study period (Figure 1). Five drought categories were applied from non-drought (0<SMADI<1) to extreme drought (SMADI>4). Note that in the application of SMADI at global scale, the results in (4) were not normalized, as done in previous research [7].

As an example, the maps of spring-summer of 2012 depicted the severity of the agricultural drought in the Indian region (Figure 1), recognized as one of the more severe recent agricultural droughts with devastating impacts on crop production and livestock [16].

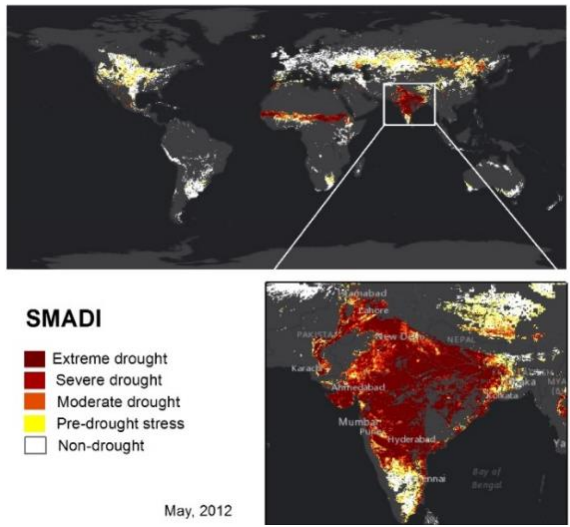

Figure 1. SMADI map from May, 2012 at global scale and at a more detailed scale covering the Indian Peninsula.

### 3.1. Comparison of SMADI with CMI and SWDI

Series of biweekly SWDI were calculated at each REMEDHUS station using their specific soil water parameters. Later, they were area-averaged and compared to the SMADI series at the network level (Figure 2). The correlation coefficient between both series was R=-0.83 (statistically significant at 99% confidence level). Interestingly, SMADI characterized 2012 and 2014 as the driest years (reaching values above 4, considered as extreme drought), whereas for SWDI all the summer periods reached the extreme drought conditions (below -10).

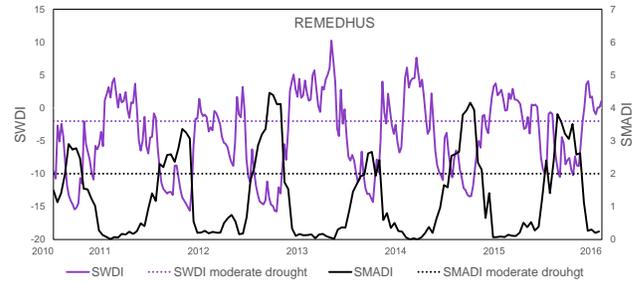

Figure 2. Temporal evolution of SWDI and SMADI the REMEDHUS area average. Thresholds of moderate drought are also showed

The comparison of CMI with SMADI at the AEMet stations (Figure 3, only the Burgos station is included) also showed good inverse correlations (Table 1). In this case, 2012 can be identified from the SMADI time series as the driest year (values higher than 4 considered extreme drought).

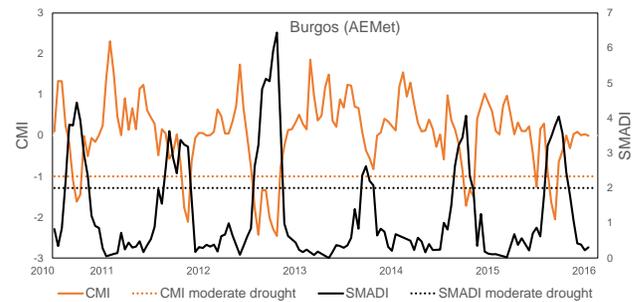

Figure 3. Temporal evolution of CMI and SMADI at the Burgos station from AEMet. Thresholds of moderate drought were also showed.

Table 1. Correlation coefficients (Pearson) between CMI and SMADI at the five AEMet stations. All cases are significant at 99% confidence level.

|   | Burgos | León | Soria | Valladolid | Zamora |
|---|---|---|---|---|---|
| R | -0.79 | -0.78 | -0.70 | -0,64 | -0.78 |

### 3.2. Comparison of SMADI with VegDRI

The comparison between VegDRI and SMADI was made on a spatial basis. Results for three particular drought events are shown in Figure 4. Correlations between maps for each biweekly period were calculated, resulting in a total of 145 correlation values ranging from R≈0 to -0.64. A persistent temporal pattern was found during the summer periods (with R better than -0.30 all of them statistically significant at 99%).

The legend of VegDRI was adapted to fit that of SMADI (Figure 4) so that the similarity of spatial patterns can be visually assessed. Note, for instance, that the severe drought in Texas and Oklahoma in 2011 [17] is consistently captured by the two indices.

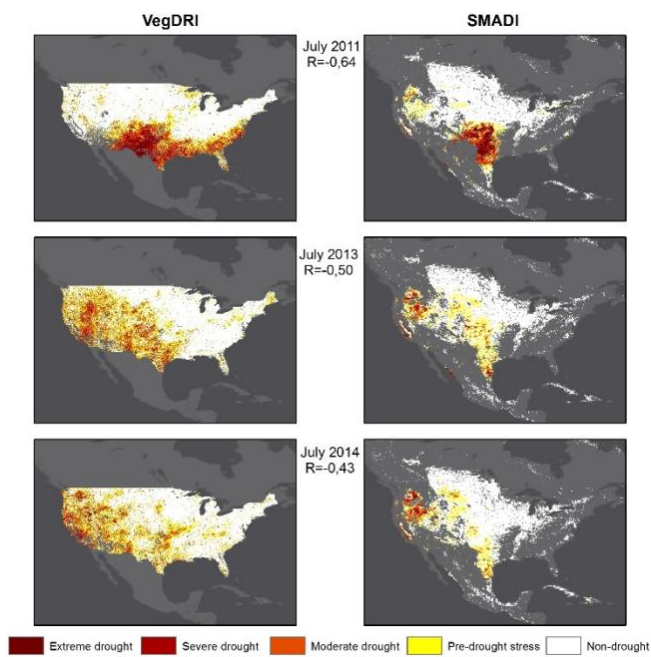

Figure 4. Maps of VegDRI and SMADI at the United States for summer 2011, 2013 and 2014.

## 4. CONCLUSIONS

The use of a unique indicator of agricultural drought at the global scale is challenging due to the variety of environmental conditions, climates and land-covers existing in the planet. Besides, the thresholds of the drought indices and thus the drought detection should be carefully selected given the number and variety of sectors that can be affected. For the first time, the Soil Moisture Agricultural Drought Index (SMADI) has been applied at the global scale. Preliminary results at several locations of the Iberian Peninsula (climatic and soil moisture networks) showed a good agreement with other agricultural drought indices (SWDI and CMI). Also, the spatial patterns of SMADI were compared with those obtained by the VegDRI index, resulting in a traceable feature that was detected during the most intense drought-affected periods in the United States.

This study shows the value of using SMADI to detect and monitor drought events along the globe using readily-available remote sensing data. Further research is needed to validate the index at additional locations distributed world-wide at different temporal and spatial scales.

*Acknowledgements*
This study was supported by the Spanish Ministry of Economy and Competitiveness (Project ESP2015-67549-C3-3-R), the Castilla y León Region Government (Project SA007U16) and the European Regional Development Fund (ERDF). The authors also acknowledge the AEMet (Spanish Meteorological Agency) for providing the climatic databases.